\documentclass{appolb}
\usepackage{enumitem}
\newbox\hdbox%
\newcount\hdrows%
\newcount\multispancount%
\newcount\ncase%
\newcount\ncols
\newcount\nrows%
\newcount\nspan%
\newcount\ntemp%
\newdimen\hdsize%
\newdimen\newhdsize%
\newdimen\parasize%
\newdimen\spreadwidth%
\newdimen\thicksize%
\newdimen\thinsize%
\newdimen\tablewidth%
\newif\ifcentertables%
\newif\ifendsize%
\newif\iffirstrow%
\newif\iftableinfo%
\newtoks\dbt%
\newtoks\hdtks%
\newtoks\savetks%
\newtoks\tableLETtokens%
\newtoks\tabletokens%
\newtoks\widthspec%
\immediate\write15{%
CP SMSG GJMSINK TEXTABLE --> TABLE MACROS V. 851121 JOB = \jobname%
}%
\tableinfotrue%
\catcode`\@=11
\def\tstrut{\vrule height3.1ex depth1.2ex width0pt}%
\def\and{\char`\&}
\def\tablerule{\noalign{\hrule height\thinsize depth0pt}}%
\def\thickrule{\noalign{\hrule height\thicksize depth0pt}}%
\def\ctr#1{\hfil\ #1\hfil}%
\tablewidth=-\maxdimen%
\spreadwidth=-\maxdimen%
\def\tabskipglue{0pt plus 1fil minus 1fil}%
\centertablestrue%
\gdef\ARGS{########}
\gdef\headerARGS{####}
\def\@mpersand{&}
{\catcode`\|=13
\gdef\letbarzero{\let|0}
\gdef\letbartab{\def|{&&}}%
\gdef\letvbbar{\let\vb|}%
}
{\catcode`\&=4
\def\ampskip{&\omit\hfil&}
\catcode`\&=13
\let&0
\xdef\letampskip{\def&{\ampskip}}%
\gdef\letnovbamp{\let\novb&\let\tab&}
}
\def\begintable{
   \begingroup%
   \catcode`\|=13\letbartab\letvbbar%
   \catcode`\&=13\letampskip\letnovbamp%
   \def\multispan##1{
      \omit \mscount##1%
      \multiply\mscount\tw@\advance\mscount\m@ne%
      \loop\ifnum\mscount>\@ne \sp@n\repeat%
   }
   \def\|{%
      &\omit\widevline&%
   }%
   \ruledtable
}
\long\def\ruledtable#1\endtable{%
   \offinterlineskip
   \tabskip 0pt
   \def\widevline{\vrule width\thicksize}
   \def\endrow{\@mpersand\omit\hfil\crnorm\@mpersand}%
   \def\crthick{\@mpersand\crnorm\thickrule\@mpersand}%
   \def\crthickneg##1{\@mpersand\crnorm\thickrule
          \noalign{{\skip0=##1\vskip-\skip0}}\@mpersand}%
   \def\crnorule{\@mpersand\crnorm\@mpersand}%
   \def\crnoruleneg##1{\@mpersand\crnorm
          \noalign{{\skip0=##1\vskip-\skip0}}\@mpersand}%
   \let\nr=\crnorule
   \def\endtable{\@mpersand\crnorm\thickrule}%
   \let\crnorm=\cr
   \edef\cr{\@mpersand\crnorm\tablerule\@mpersand}%
   \def\crneg##1{\@mpersand\crnorm\tablerule
          \noalign{{\skip0=##1\vskip-\skip0}}\@mpersand}%
   \let\ctneg=\crthickneg
   \let\nrneg=\crnoruleneg
   \the\tableLETtokens
   \tabletokens={&#1}
   \countROWS\tabletokens\into\nrows%
   \countCOLS\tabletokens\into\ncols%
   \advance\ncols by -1%
   \divide\ncols by 2%
   \advance\nrows by 1%
   \iftableinfo %
      \immediate\write16{[Nrows=\the\nrows, Ncols=\the\ncols]}%
   \fi%
   \ifcentertables
      \ifhmode \par\fi
      \hbox to \hsize{
      \hss
   \else %
      \hbox{%
   \fi
      \vbox{%
         \makePREAMBLE{\the\ncols}
         \edef\next{\preamble}
         \let\preamble=\next
         \makeTABLE{\preamble}{\tabletokens}
      }
      \ifcentertables \hss}\else }\fi
   \endgroup
   \tablewidth=-\maxdimen
   \spreadwidth=-\maxdimen
}
\def\makeTABLE#1#2{
   {
   \let\ifmath0
   \let\header0
   \let\multispan0
   \ncase=0%
   \ifdim\tablewidth>-\maxdimen \ncase=1\fi%
   \ifdim\spreadwidth>-\maxdimen \ncase=2\fi%
   \relax
   \ifcase\ncase %
      \widthspec={}%
   \or %
      \widthspec=\expandafter{\expandafter t\expandafter o%
                 \the\tablewidth}%
   \else %
      \widthspec=\expandafter{\expandafter s\expandafter p\expandafter r%
                 \expandafter e\expandafter a\expandafter d%
                 \the\spreadwidth}%
   \fi %
   \xdef\next{
      \halign\the\widthspec{%
      #1
      \noalign{\hrule height\thicksize depth0pt}
      \the#2\endtable
      }
   }
   }
   \next
}
\def\makePREAMBLE#1{
   \ncols=#1
   \begingroup
   \let\ARGS=0
   \edef\xtp{\widevline\ARGS\tabskip\tabskipglue%
   &\ctr{\ARGS}\tstrut}
   \advance\ncols by -1
   \loop
      \ifnum\ncols>0 %
      \advance\ncols by -1%
      \edef\xtp{\xtp&\vrule width\thinsize\ARGS&\ctr{\ARGS}}%
   \repeat
   \xdef\preamble{\xtp&\widevline\ARGS\tabskip0pt%
   \crnorm}
   \endgroup
}
\def\countROWS#1\into#2{
   \let\countREGISTER=#2%
   \countREGISTER=0%
   \expandafter\ROWcount\the#1\endcount%
}%
\def\ROWcount{%
   \afterassignment\subROWcount\let\next= %
}%
\def\subROWcount{%
   \ifx\next\endcount %
      \let\next=\relax%
   \else%
      \ncase=0%
      \ifx\next\cr %
         \global\advance\countREGISTER by 1%
         \ncase=0%
      \fi%
      \ifx\next\endrow %
         \global\advance\countREGISTER by 1%
         \ncase=0%
      \fi%
      \ifx\next\crthick %
         \global\advance\countREGISTER by 1%
         \ncase=0%
      \fi%
      \ifx\next\crnorule %
         \global\advance\countREGISTER by 1%
         \ncase=0%
      \fi%
      \ifx\next\crthickneg %
         \global\advance\countREGISTER by 1%
         \ncase=0%
      \fi%
      \ifx\next\crnoruleneg %
         \global\advance\countREGISTER by 1%
         \ncase=0%
      \fi%
      \ifx\next\crneg %
         \global\advance\countREGISTER by 1%
         \ncase=0%
      \fi%
      \ifx\next\header %
         \ncase=1%
      \fi%
      \relax%
      \ifcase\ncase %
         \let\next\ROWcount%
      \or %
         \let\next\argROWskip%
      \else %
      \fi%
   \fi%
   \next%
}
\def\counthdROWS#1\into#2{%
\dvr{10}%
   \let\countREGISTER=#2%
   \countREGISTER=0%
\dvr{11}%
\dvr{13}%
   \expandafter\hdROWcount\the#1\endcount%
\dvr{12}%
}%
\def\hdROWcount{%
   \afterassignment\subhdROWcount\let\next= %
}%
\def\subhdROWcount{%
   \ifx\next\endcount %
      \let\next=\relax%
   \else%
      \ncase=0%
      \ifx\next\cr %
         \global\advance\countREGISTER by 1%
         \ncase=0%
      \fi%
      \ifx\next\endrow %
         \global\advance\countREGISTER by 1%
         \ncase=0%
      \fi%
      \ifx\next\crthick %
         \global\advance\countREGISTER by 1%
         \ncase=0%
      \fi%
      \ifx\next\crnorule %
         \global\advance\countREGISTER by 1%
         \ncase=0%
      \fi%
      \ifx\next\header %
         \ncase=1%
      \fi%
\relax%
      \ifcase\ncase %
         \let\next\hdROWcount%
      \or%
         \let\next\arghdROWskip%
      \else %
      \fi%
   \fi%
   \next%
}%
{\catcode`\|=13\letbartab
\gdef\countCOLS#1\into#2{%
   \let\countREGISTER=#2%
   \global\countREGISTER=0%
   \global\multispancount=0%
   \global\firstrowtrue
   \expandafter\COLcount\the#1\endcount%
   \global\advance\countREGISTER by 3%
   \global\advance\countREGISTER by -\multispancount
}%
\gdef\COLcount{%
   \afterassignment\subCOLcount\let\next= %
}%
{\catcode`\&=13%
\gdef\subCOLcount{%
   \ifx\next\endcount %
      \let\next=\relax%
   \else%
      \ncase=0%
      \iffirstrow
         \ifx\next& %
            \global\advance\countREGISTER by 2%
            \ncase=0%
         \fi%
         \ifx\next\span %
            \global\advance\countREGISTER by 1%
            \ncase=0%
         \fi%
         \ifx\next| %
            \global\advance\countREGISTER by 2%
            \ncase=0%
         \fi
         \ifx\next\|
            \global\advance\countREGISTER by 2%
            \ncase=0%
         \fi
         \ifx\next\multispan
            \ncase=1%
            \global\advance\multispancount by 1%
         \fi
         \ifx\next\header
            \ncase=2%
         \fi
         \ifx\next\cr       \global\firstrowfalse \fi
         \ifx\next\endrow   \global\firstrowfalse \fi
         \ifx\next\crthick  \global\firstrowfalse \fi
         \ifx\next\crnorule \global\firstrowfalse \fi
         \ifx\next\crnoruleneg \global\firstrowfalse \fi
         \ifx\next\crthickneg  \global\firstrowfalse \fi
         \ifx\next\crneg       \global\firstrowfalse \fi
      \fi
\relax
      \ifcase\ncase %
         \let\next\COLcount%
      \or %
         \let\next\spancount%
      \or %
         \let\next\argCOLskip%
      \else %
      \fi %
   \fi%
   \next%
}%
\gdef\argROWskip#1{%
   \let\next\ROWcount \next%
}
\gdef\arghdROWskip#1{%
   \let\next\ROWcount \next%
}
\gdef\argCOLskip#1{%
   \let\next\COLcount \next%
}
}
}
\def\spancount#1{
   \nspan=#1\multiply\nspan by 2\advance\nspan by -1%
   \global\advance \countREGISTER by \nspan
   \let\next\COLcount \next}%
\def\dvr#1{\relax}%
\def\header#1{%
\dvr{1}{\let\cr=\@mpersand%
\hdtks={#1}%
\counthdROWS\hdtks\into\hdrows%
\advance\hdrows by 1%
\ifnum\hdrows=0 \hdrows=1 \fi%
\dvr{5}\makehdPREAMBLE{\the\hdrows}%
\dvr{6}\getHDdimen{#1}%
{\parindent=0pt\hsize=\hdsize{\let\ifmath0%
\xdef\next{\valign{\headerpreamble #1\crnorm}}}\dvr{7}\next\dvr{8}%
}%
}\dvr{2}}
\def\makehdPREAMBLE#1{
\dvr{3}%
\hdrows=#1
{
\let\headerARGS=0%
\let\cr=\crnorm%
\edef\xtp{\vfil\hfil\hbox{\headerARGS}\hfil\vfil}%
\advance\hdrows by -1
\loop
\ifnum\hdrows>0%
\advance\hdrows by -1%
\edef\xtp{\xtp&\vfil\hfil\hbox{\headerARGS}\hfil\vfil}%
\repeat%
\xdef\headerpreamble{\xtp\crcr}%
}
\dvr{4}}
\def\getHDdimen#1{%
\hdsize=0pt%
\getsize#1\cr\end\cr%
}
\def\getsize#1\cr{%
\endsizefalse\savetks={#1}%
\expandafter\lookend\the\savetks\cr%
\relax \ifendsize \let\next\relax \else%
\setbox\hdbox=\hbox{#1}\newhdsize=1.0\wd\hdbox%
\ifdim\newhdsize>\hdsize \hdsize=\newhdsize \fi%
\let\next\getsize \fi%
\next%
}%
\def\lookend{\afterassignment\sublookend\let\looknext= }%
\def\sublookend{\relax%
\ifx\looknext\cr %
\let\looknext\relax \else %
   \relax
   \ifx\looknext\end \global\endsizetrue \fi%
   \let\looknext=\lookend%
    \fi \looknext%
}%
\def\tablelet#1{%
   \tableLETtokens=\expandafter{\the\tableLETtokens #1}%
}%
\catcode`\@=12
\def\gev{\, \mbox{GeV}}
\def\tev{\, \mbox{TeV}}
\def\mev{\, \mbox{MeV}}
\def\KL{K\"all\'en-Lehmann\,}
\def\del{\partial}
\def\TOU{theory of the universe\,}
\def\TOE{theory of everything\,}
\def\CS{Chern-Simons\,}
\def\Journal#1#2#3#4{{#1} {\bf #2}, #3 (#4)}

\def\NPB{{\em Nucl. Phys.} B}
\def\PLB{{\em Phys. Lett.}  B}
\def\PRL{\em Phys. Rev. Lett.}
\def\PRD{{\em Phys. Rev.} D}

\def\beq{\begin{equation}}
\def\eeq{\end{equation}}
\def\bea{\begin{eqnarray}}
\def\eea{\end{eqnarray}}

\begin{document}

\preprint{3333}

\title{The Principle of Global Relativity
}
\author{Jochum~Johan~van~der~Bij
\address{
Insitut f\"ur Physik, Albert-Ludwigs Universit\"at Freiburg \\
H. Herderstr. 3, 79104 Freiburg i.B., Deutschland
}}

\maketitle

\begin{center}
{ \sf  In memoriam Prof. M.J.G. Veltman\\
 for Tini's 90th birthday~[m1]\footnote{Citations [m1], [m2], etc. refer to the paragraphs in the last, memoir Section.}.}
\end{center}
\begin{abstract}
We describe the non-minimal Standard Model, consisting of minimalistic
extensions of the Standard Model, which for all we know is the theory of the universe,
able to describe all of the universe from the beginning of time.
Extensions discussed are an extra neutrino and a new Higgs model.
We introduce the principle of global relativity and discuss how the theory can be largely
derived from this principle. One is led to the  unification of forces into SU(5)
and a form of dark matter.
 We discuss the limitations of the theory, showing that it is not the theory of everything.
However we argue that it is the only part, that is within conceivable reach of physical experiment or 
astronomical observation. 
It is argued, that at the Planck scale the universe is effectively three-dimensional. 
\end{abstract}


\section{Introduction - theory of the universe}

Underlying the endeavour of fundamental physics is the conviction, that
nature can be described by unbreakable fundamental laws, that take the form of
mathematical equations. To find out what these laws are, one uses the so-called
scientific method. This method consists of performing experiments, guided by theoretical ideas
and inventing theoretical ideas, guided by experiment. This all follows common sense.
This must be so. After all it is easy to do senseless experiments or to invent crazy theories.
In certain popular presentations and books on the "philosophy" of science this is not always clear.
The scientific method has been enormously successful recently in the discovery of the Higgs boson
and  of gravitational waves. The first follows from a combination of the special theory
of relativity and quantum mechanics, the second from the combination of special relativity 
and gravity. So one gains confidence, that one may have found at least a large part of the laws of nature.
In this paper I will argue, that we know enough by now to say that we know the laws of nature sufficiently to
explain all of the universe that we can see. We know, so to say, the theory of the universe.
Assuming this to be true, we can try to move up a level and ask whether we can find a reason why
precisely these laws are the right ones. This has been tried rather unsuccessfully in the past,
where people tried to find a theory of everything. Here we will make a new try and argue that 
we may have come halfway.

A theory of the universe should be able to describe at least in principle how the universe
developed from the beginning of time to its present form. 
Since the time I started studying physics in Utrecht in 1974 we have come a long way in this direction.
There are actually four questions that should be answered:\\
\noindent
1) why are the experiments so consistent with the Standard Model?\\ 
2) how did the baryon number of the universe arise?\\
3) how did the universe come into existence from inflation?\\
4) what is the dark matter?
\\

There are many proposals in the literature, that attempt to explain these
four points. These can take various forms of complexity, involving 
sophisticated concepts, like string theory or supersymmetry, that will
predict many more things than what we see in nature, These are maximalistic models
that tend to predict phenomena that simply are not there, in conflict with point 1.
The point of view we take here is, that already the simplest possible extensions,
minimalistic extensions, of the Standard Model are sufficient.
By minimalistic extensions I mean extensions that do not change the fundamental chiral
gauge structure of the Standard Model. Examples are non-chiral fermions, inert multiplets,
St\"uckelberg-like vector bosons and singlets in general. I call this class of models
the Non-Minimal Standard Model (NMSM). In discussions one often hears about the Standard Model (SM), the minimal
Supersymmetric Standard Model (MSSM) and the non-minimal Supersymmetric Standard Model (NMSSM).
As the SM is not complete and the MSSM and the NMSSM have severe phenomenological
difficulties, the NMSM may be the truth~[m2]. The favorite explanation for the baryon number of
the universe nowadays is leptogenesis, whereby heavy neutrinos generate a lepton number in
the early universe, that subsequently gets transformed into baryons through
sphaleron processes   \cite{murayama,klinkhamer,kuzmin}. In section 2 we will study precision
experiments to see if there is evidence for the existence of 
extra neutrinos. One of the most popular theories of inflation is Higgs-inflation,
where the Higgs boson acts as the inflaton  \cite{shaposh}. In section 3 we will extensively
study Higgs physics and construct minimalistic extensions that can work for inflation as well.
In section 5 we introduce a new principle, with which we address the question why the standard 
model should be the low energy theory. We are led to a model with a unification 
into SU(5), that implies a fairly unique answer to question 4.
Therefore, we arrive altogether at a consistent view of the world, whereby at low energies 
indeed a form of the NMSM describes nature in a way sufficient  to describe the development of
the universe as far as we can study it. Beyond that we found a reason, why this should be the case.
In the final section we discuss the limitations of the theory and make some speculations about a theory of everything. 

\section{Precision predictions}
The core of Tini's work is the calculation of radiative corrections
due to the weak interactions. Without the possibility of performing
such calculations, one cannot speak of a theory. This was the situation before Tini entered
the field. The situation now is different. Calculations due to weakly interacting particles
have in the meantime been done even to the three-loop level in certain cases.
This is of course only possible in a renormalizable theory.
Such calculations would not have been possible without the development of 
computer algebra programs, of which Tini's SCHOONSCHIP was the first program
that could handle the large expressions appearing in quantum field theory
calculations. Comparing precise calculations with precise measurements
can under circumstances even give an indication of the existence of
new particles. I will give two examples of this.

\subsection{$\rho$ parameter}

One of the quantities that is sensitive to effects of
heavy particles is the so called $\rho$ parameter   \cite{ross}.
That is the ratio between the neutral and the charged Fermi constant $\rho = G_F^0/G_F^+$.
A peculiar feature of the Standard Model in comparison with other gauge models
is that at the tree level $\rho$ is equal to one.  
This is due to an accidental O(4) symmetry of
the Higgs sector, that is larger than the symmetry SU(2) $\times$ U(1) of the Standard Model as a whole.
It receives radiative corrections from mass splittings within fermion multiplets and through 
hypercharge couplings. In particular the effects of a heavy top quark can become quite large.
This is due to the fact that the mass of the top quark is proportional to its
Yukawa coupling, which becomes large, so that we have strong interactions present.   
As a consequence even at the one-loop level effects growing like $m_t^2$ are present \cite{tinitop}.
Two-loop effects grow like $m_t^4$ \cite{hoogeveen} and three-loop effects like $m_t^6$ \cite{chety}.
The results can be summarized in an effective Lagrangian \cite{steger}.
However, just like in the Fermi theory, this Lagrangian cannot be used for loop calculations,
since the loop effects can only be determined in the underlying renormalizable theory.
This is a rather common problem for effective Lagrangians.
Similar large top mass effects appear in couplings of the Z-boson to bottom quarks.
In principle, having precise enough data at low energy, one could have predicted the top mass.
This is not quite how things happened historically. There were some indications of a heavy top quark
in the bottom data, but not as large as $170\gev$ where it was found at Fermilab.
The LEP data only agreed with the Fermilab result after a reanalysis of the data.

\subsection{Lepton non-universality}
With the mass of the top quark known, one can try to predict the mass of the Higgs boson from
the precision measurements. For this the so-called blue-band plot was invented, which  listed the
$(\Delta \chi)^2$ from the best fit for the Higgs mass. But this was always controversial,
since the overall $\chi^2$ was large and so the fit was not good. The problem lies with
the bottom quark asymmetry in the decay of the Z-boson.
 Leaving out this measurement from the
data, arguments for the existence of new physics were made \cite{channewp, lorca}. But, because the Higgs mass 
was unknown, these attempts were inconclusive. After the discovery of the Higgs boson things changed.

The importance of the results at the LHC is that the Higgs boson mass has been determined
and that no new particles, carrying weak charges appear to exist. This implies that we can 
compare theory predictions with data to a much higher level of precision than before. Precise predictions
in the theory are sensitive to radiative corrections, dependent on the Higgs boson mass. 
Before the discovery of the Higgs boson, the data were used to constrain the range of the mass for the Higgs boson.
Now that all parameters of the model are known the theory predictions are essentially exact, so one can 
look for much smaller deviations than before. To look for possible deviations in the precision data,
we consider a model with $n$ neutral sterile fermions, that only mix with the neutrinos of the Standard Model.
Such particles can play a role in cosmology, i.e. in leptogenesis or as dark matter candidates.
The consequence is that the Pontecorvo-Maki-Nakagawa-Sakata ($PMNS$) matrix is part of a more general mixing matrix.

Taking into account the Standard Model neutrinos and the extra neutrinos we find that
the mass eigenstates $( \nu_1 \cdots  \nu_{3+n} )$
and flavour basis ($\alpha= e,\mu,\tau$): $\{\nu_i=\nu_{L_\alpha},\,N_n\}$ are connected by a unitary $(3+n)\times(3+n)$ matrix:\\

\begin{displaymath}
\left(\begin{array}{c} \nu_1 \\ \vdots \\ \nu_{3+n} \end{array}\right) =  \left(\begin{array}{cc} PMNS & {\cal W} \\ {\cal W}^\dagger & {\cal V} \end{array}\right) \left(\begin{array}{c} \nu_{L_e} \\ \vdots \\ N_n\end{array}\right)\,.
\end{displaymath}
As a consequence the
$PMNS$ matrix, being a  submatrix, is not necessarily unitary.
We describe the deficit from unitarity by the $\epsilon$ parameters:

\begin{displaymath}
\epsilon_\alpha = \sum_{i>3} |{\cal U}_{\alpha i}|^2 = 1- \sum_{\beta= e, \mu, \tau }|{\cal U}_{\alpha \beta}|^2\,.
\end{displaymath}

As a consequence low energy parameters are affected by the $\epsilon$ parameters.
For example the Fermi constant in muon-decay is modified by the following relation:\\ 

\begin{displaymath}
G_\mu^2 = G_F^2 (1-\epsilon_e)(1-\epsilon_\mu)\,,
\label{eq-Gmu}
\end{displaymath}
with $G_\mu$ the Fermi constant  measured in muon decay, and $G_F$  the Fermi parameter, derived from the Standard Model theory
without $\epsilon$ parameters.

Other corrections appear in meson-decays and in precision measurements at LEP; we are therefore in the
lucky position that we can combine low-energy and high-energy (LEP) measurements. 
When we do this, we find that the most precise data cannot be well fitted to the model  \cite{basso,antusch}, even allowing
for the presence of the $\epsilon$ parameters. The origin of the problem was tracked to a single measurement,
namely the forward-backward asymmetry of bottom quarks at LEP. This measurement would lead to a  large and unphysical negative value
for $\epsilon_e$. The other measurements are in good agreement with each other, leading to a value
$\epsilon_e \approx 2.10^{-3}$, excluding $\epsilon_e =0$ at the 2-3~$\sigma$ level. 
A number of experiments that can test this result are underway, new measurements of $\sin^2_{eff}(\theta_W)$, an improvement
on $M_W$, meson decays in $b$ and $\tau$ factories, the ratio $W\rightarrow e$/$W\rightarrow \mu$ and a precise
lattice evaluation of $f_{\pi}$. In combination these could lead to a 5$\sigma$ discovery.

\section{Higgs physics}

\subsection{heavy Higgs}
With the construction of the Standard Model, it became clear, that in some way one would have to look
for the Higgs boson itself in colliders~[m3]. The problem was, that there was no way to tell what its
mass was. Therefore it was somewhat unclear, what sort of collider one needed.
An upper mass for the Higgs boson of about $1\tev$ was found in  \cite{leequitha}, based on tree 
level unitarity violation, indicating the start of possible strong interaction phenomena
like multiple vectorboson production. Actually the more relevant scale is probably $4\pi v \approx 3\tev$, where interaction
effects become large. Many people at the time thought the Higgs boson would not really be fundamental,
but more of an effective description coming from a strongly interacting underlying theory,
somewhat analogous to QCD. Of course a simple scalar was also possible, but in general people were somewhat
suspicious of the mechanism~[m4]. As a conclusion it was decided that one would have to build
a large energy hadron collider, SSC or LHC in order to cover the full range of possibilities;
in the end only the LHC was built.
The situation was summarized in the so-called no-lose scenario:
building such a machine would either lead to the discovery of the Higgs boson or to the finding
of new strong interactions \cite{nolose}. Of course the most important point of such scenarios
is how to avoid them, which was indeed possible but not so easy.
With strong enough mixing, the HEIDI-models to be described below, would have made the Higgs
boson undetectable at the LHC. It is somewhat debatable, whether one should call finding the Higgs boson a discovery. 
After all, barring miraculous fine-tunings in other theories, the Standard Model was the only
quantum mechanically consistent theory of the weak interaction, so one is actually only
confirming quantum mechanics. But no one seriously doubts the validity of quantum mechanics.
Not finding the Higgs boson however would have been a real discovery.  

Complementary to large effects at high energy, there are large effects at low energies through radiative corrections.
So a large Higgs mass, that leads to large cross sections at high energy, also will give rise
to effects that rise with the Higgs mass in loop effects. This was the subject of the
first discussion I had with Tini in 1979. He asked me  how one could see a strongly interacting Higgs. This was in 
first instance meant in radiative corrections \cite{screen}.  Within the Standard Model that would mean a heavy Higgs.
Though in the end the Higgs boson is light, the question has a number of interesting aspects,
for instance related to the Landau pole and to the decoupling theorem. From the technical point of view having a heavy Higgs
makes higher-loop graphs simpler. Starting with an arbitrary Higgs mass, would have been prohibitively
difficult given the technical possibilities of the time. Even now two-loop graphs with arbitrary masses are still
at the edge of technique; examples are  \cite{gudrun,bonciani}. For massless fields one can go to much higher order \cite{peter}.
This is only possible due to the method of dimensional regularization \cite{dimreg}.

The radiative corrections dependent on the Higgs mass form a special case, different from the effects
of other heavy particles. Normally particles that are heavy decouple from the theory, since their
effects are suppressed by the square of the mass. Typical is for instance the muon, that does not affect atomic physics.
The other extreme is the top quark, where the mass is proportional to a coupling, and so one gets strong effects
for a heavy mass. The Higgs case is in between. The theory without a Higgs boson is close to renormalizable 
and at the one-loop level one only gets logarithmically divergent effects. The theory without a Higgs particle
can be seen as the limit of infinite Higgs mass, keeping the vacuum expectation value of the Higgs field
constant. This means taking the self-coupling to infinity. This can be done at the tree level without problem;
at the one-loop level one gets logarithmically divergent corrections. But because of the strong 
coupling this is not sufficient for an ultimate conclusion and one needs a more detailed
analysis. As a starting point one can calculate two-loop effects for the $\rho$ parameter \cite{vdbij83}~[m5,m6].
One finds effects growing like $m_H^2$. Also other quantities like mass-shifts of the vectorbosons
or anomalous self-couplings behave in the same way; this is also true for the subsequent calculations.
A first bound on the Higgs self-coupling was found in  \cite{vdbij85}. A new aspect arises in the
four-vector boson couplings. Here also Higgs-reducible diagrams contribute  \cite{jikia}.
In these calculations special numbers like $Cl(\pi/3)$ appear, that play a role in number theory, having connections
with elliptic functions \cite{blumlein}. Altogether the effects were rather small. So the next step was to
look at the Higgs propagator itself at the two-loop level \cite{adrian}.  Here larger effects appeared.
A logical next step is then to take a non-perturbative approach by making a $1/N$ expansion of the theory,
whereby $N=4$, given the fact that the Higgs sector is an $O(4)$ model.  
The lowest order is trivial to calculate,
but not very accurate. The next order was calculated in  \cite{binoth}. This was highly non-trivial,
introducing new features like tachyon subtraction. The approach also goes by the name of renormalon physics
 or causal perturbation theory and is more familiar in QCD. The two-loop and the second order $1/N$ result
agreed rather well, giving the following picture. The Higgs becomes very wide with a mass peak at about $1\tev$ \cite{binothlet}.
However it is the width that is the measure of the coupling strength. The one-loop corrections to the Higgs propagator
are small however. That also explains why the two-loop corrections to the low energy parameters are small.
The qualitative reason is that the s-dependent width of the Higgs first appears at the two-loop order in the propagator.
So it became necessary to calculate the three-loop correction to the $\rho$ parameter \cite{radja}.
This was indeed larger than the two-loop one. Therefore one finally has a consistent picture of
a heavy Higgs. The Higgs gets wide, which means that the \KL spectral density of the Higgs field has a high mass
component. In the radiative corrections one then practically has a one-loop correction with a Higgs mass
replaced by the spectral density \cite{akhoury}. This shows that a strongly interacting Higgs will lead to large effects
in the radiative corrections. The strong interactions enhance the one-loop effects and cannot cancel them.
One loophole is still left. Maybe cross exchange between the Higgs and the vectorboson in the loop 
could become large, leading to bound states between vectorbosons and Higgs particles.
This was studied in \cite{dilcherthesis}. Qualitatively the situation is clear. In order to compensate for the one-loop 
$log(m_H)$ effects one would need low lying bound states mixing with the vectorbosons, which one should have seen.
The conclusion is therefore clear. If precision measurements based on a one-loop analysis indicate that the Higgs is light,
it must be light. Strong interactions cannot compensate the one-loop effects.

\subsection{Effective field theories}
With the discovery of the Higgs particle the impression exists that high-energy physics
is finished and that there is no real reason to build a new accelerator.
However it is argued that the measurement of the Higgs self-coupling is necessary to further establish the Standard Model.
This is one of the main arguments for extending the LHC to a higher luminosity, the HL-LHC.
However the argument is somewhat weak, because the precision with which one can measure the three-Higgs
self-coupling is rather small, the four-Higgs self-coupling is completely out of reach.
So this is a weak test, in particular since it is difficult to make models that would generate 
a large Higgs self-coupling. Anomalous Higgs self-couplings can be generated by loop effects,
for instance through the exchange of singlet scalars. However these are small and effects would have shown up 
in other precision experiments before. 
Therefore changes in the Higgs sector tend to be described by so-called effective field theories (EFT's) \cite{eft}.
What one does here is to add higher-dimensional point-like interactions, that can parametrize 
deviations from the Standard Model couplings. The term is a bit of a misnomer. Traditionally one describes
couplings by writing general Lorentz-invariant amplitudes, also allowing for form factors.
In an EFT form factors are made pointlike and one compares with the data at the tree-level.
This is not a very consistent procedure, since it is not possible to do loop corrections in a way that can
be compared with electroweak precision measurements. In order to really have a 
quantum mechanical theory, one needs to impose the Schwinger-Dyson equations, that have to be made finite first. 
In order to do this one has to change the propagators too, generating a cut-off to the theory.
In general things become quite complicated and one loses effectively any predictivity. 
An example of such an analysis for anomalous vector boson couplings is given in  \cite{kastening}.
Another example, that we discussed before, is the Standard Model without a top quark. 

Moreover the energy range of the LHC is very large and one sees no deviations from the Standard Model.
Therefore the EFT's effects should be too small to be measured. The approach actually violates 
Veltman's theorem, which says that a low energy theory must be either renormalizable or strongly interacting.
 This means one should work with a renormalizable theory from the start,
which is the reason Veltman and 't Hooft received a Nobel prize in the first place.
Nonetheless not all hope is lost. At least for a limited class of EFT's it is possible to show
that they are the limit of renormalizable models, where some couplings become infinite.
At the tree level this can be done, but at the quantum level one must take the original model.
The situation is very similar to taking the infinite Higgs-mass limit within the Standard Model.
EFT's where such a construction is not possible most likely cannot correspond to real physics;
these are only a parametrization.

\subsection{The Hill model}
In order to test how good a theory like the Standard Model is, it is good practice to compare
with a model that is different in the most minimal way. 
The simplest  possible  renormalizable extension of the Standard Model is
actually the Hill model \cite{hill}, having only two extra parameters~[m7].
The Hill model is described by the following Lagrangian:
\begin{eqnarray*}
{\cal L} =&& -\frac{1}{2}(D_{\mu} \Phi)^{\dagger}(D_{\mu} \Phi) 
-\frac {1}{2}(\partial_{\mu} H)^2 \\
&&- \frac {\lambda_0}{8} 
(\Phi^{\dagger} \Phi -f_0^2)^2  
-\frac {\lambda_1}{8}(2f_1 H-\Phi^{\dagger}\Phi)^2 .
\end{eqnarray*}
Working in the unitary gauge one writes $\Phi^{\dagger}=(\sigma,0)$,
where the $\sigma$-field is the physical Standard Model Higgs field.
Both the SM Higgs field $\sigma$ and the Hill field $H$ receive vacuum expectation
values and one ends up with a two-by-two mass matrix to diagonalize, thereby
ending with two masses $m_-$ and $m_+$ and a mixing angle $\alpha$. There are two
equivalent ways to describe this situation. One is to say that one has two Higgs
fields with reduced couplings $g$ to Standard Model particles:
\begin{displaymath}
g_-= g_{SM} \cos(\alpha), \qquad g_+= g_{SM} \sin(\alpha) .
\end{displaymath} 
The Standard Model would correspond to $\alpha=0$ with the light Higgs
the Standard Model Higgs.
The other way, which has some practical advantages is not to diagonalize
the propagator, but simply keep the $\sigma - \sigma$ propagator explicitely.
One simply replaces the Standard Model Higgs propagator,
in all calculations of experimental cross section,
 by:
\begin{displaymath}
D_{\sigma \sigma} (k^2) = \cos^2(\alpha)/(k^2 + m_-^2) + \sin^2(\alpha)/(k^2 + m_+^2).
\end{displaymath}

The generalization to an arbitrary set of fields $H_k$ is straightforward, one 
simply replaces the singlet-doublet interaction term by:
\begin{displaymath}
L_{H \Phi}= - \sum \frac {\lambda_k}{8}(2f_k H_k-\Phi^{\dagger}\Phi)^2 . 
\end{displaymath}
For a finite number of fields $H_k$ no essentially new aspects appear,
however dividing the Higgs signal over even a small number of peaks,
can make the study of the Higgs field at the LHC  somewhat challenging.
Having an infinite number of Higgs fields one can also make a continuum \cite{dilcher,pulice}.
A mini-review of this type of models is given in  \cite{jochum}.

\subsection{A new Higgs model}
Here we will derive the Hill model backwards, starting from an effective field theory
and show some extra possibilities.
We start with the Standard Model and add the following effective terms to the Lagrangian

\begin{displaymath}
{\cal L}_{eff} =
 - \frac{1}{2 M^2} \del_{\mu} (\Phi^{\dagger}\Phi)\del_{\mu} (\Phi^{\dagger}\Phi) 
 - \frac{\lambda_3}{6 M^2} (\Phi^{\dagger}\Phi)^3 
 - \frac{\lambda_4}{24 M^4} (\Phi^{\dagger}\Phi)^4
\end{displaymath}

Here $\Phi$ is the ordinary Higgs field.
Now we introduce a new composite field $H$ through the formula:
\begin{displaymath}
M H = \Phi^{\dagger}\Phi
\end{displaymath}
In its form the Lagrangian now is a simple singlet-doublet model, however with the
constraint above. In the Lagrangian that would correspond to a delta function
for the fields. In quantum field theory this is represented by a steep 
potential, so we get an extra term in the Lagrangian:
\begin{displaymath}
\delta(M H -\Phi^{\dagger}\Phi) \rightarrow
\lim_{\lambda_{\delta} \to \infty} \lambda_{\delta}\,(M H -\Phi^{\dagger}\Phi)^2
\end{displaymath}
If we now take $\lambda_{\delta}$ finite, we have an ordinary singlet-doublet model,
so we see that the EFT is a singular limit of a renormalizable theory, in which radiative
corrections can be calculated. For the analysis of the data one should of course 
use the renormalizable theory. In the case $\lambda_3=\lambda_4=0$ one gets back the Hill
model.

An interesting \KL spectral density can be generated by assuming that the Hill field $H$
moves in more than four dimensions \cite{dilcher,pulice}, which can be taken to be infinite and flat.
We call such models HEIDI models, because of the german pronunciation of high-D(imensional).
In this case one is led to the following propagator,

\begin{eqnarray*}
\label{higgsprop4}
D_{\sigma \sigma}(q^2)=\Bigg(q^2 +M^2-\frac{\mu^{8-d}}{(q^2+m^2)^{\frac{6-d}{2}} \pm \nu^{6-d}} \Bigg)^{-1}.
\end{eqnarray*}
In this expression $d$ is the number of dimensions which should satisfy $d\leq 6$, in order to insure
renormalizability. Actually $d$ does not necessarily have to be integer to have a proper propagator.
The parameter $\mu$ describes the mixing of the higher-dimensional Hill field with
the Standard Model Higgs; indeed putting $\mu=0$ one gets the ordinary Higgs propagator with Higgs boson mass $M$.
The parameter $m$ is a higher-dimensional mass term and the parameter $\nu$ describes the mixing between higher
dimensional modes. Depending on the parameters this propagator describes zero, one or two peaks plus a continuum.
The continuum would correspond to a part of the Higgs field moving away in the extra dimensions; experimentally this would
be interpreted as an invisible decay. The HEIDI models can also stabilize the Higgs-potential.
For example one could have a 90\% Standard Model Higgs at $125\gev$, a 5\% Standard Model like Higgs at $142\gev$ and a 
5\% invisible continuum with an average Higgs mass-squared around $(180\gev)^2$. This would lead to a flat Higgs potential at the Planck mass.
Therefore it is important to measure the properties of the $125\gev$ Higgs boson as precisely as possible, in particular the overall
cross-section normalized to the Standard Model is of interest.  One should also look for further Standard Model-like Higgs bosons.
In these models, the branching ratios to Standard Model particles are 
the same as in the Standard Model for a Higgs boson with the same mass. The invisible continuum might be hard to see.

In this analysis is was assumed that $\lambda_3=0$. For renormalizability this is not
necessary. Up to six dimensions a $H^3$ self-interaction of the Hill field stays renormalizable,
but the tree-level potential becomes unbounded from below. This may actually not be a problem in
perturbation theory. After all, also the quantum potential of the Higgs field is possibly
unstable. The presence ot the $H^3$ self-interactions will after diagonalization affect
the Higgs triple-coupling as well. This is therefore a new and renormalizable model, that 
in principle allows for large deviations in the self-coupling of the Higgs.
Maybe these are large enough to be measured at the HL-LHC.
A further study is needed.

\section{Colliders}
Having constructed the HEIDI models, it is now clear that one cannot claim
to have established the Standard Model in the Higgs sector, without having
measured the complete \KL spectral density of the Higgs-propagator.
So far the limits on this model are quite weak. Even if one measures the ratio 
of the Standard Model cross-section and branching ratios within 10\%, still 10\%
of the spectral density could be hidden elsewhere. For the coupling constants
one would have to take the square root, so one gets about a 0.3 limit at the 
coupling constant level, which is no precision at all. The term precision Higgs physics
as usually used is therefore a misnomer. CERN has basically decided
to build the HL-LHC, which will not improve the situation much at all.
Therefore, one will need a lepton collider in order to get much better information on the Higgs sector.
There are actually three reasons to build a lepton collider~[m8].
First there is the need to redo the precision measurements at the Z-pole, in order
to clarify the discrepancies in the LEP-data, that we described in the section on neutrinos.
Secondly, there is a $2.3\sigma\, 10\%$ Higgs-like signal at $98\gev$, that one should check. A hadron machine cannot do this, 
since the branching ratio into photons is too small. Thirdly one should scan the Higgs spectrum,
looking among others for the non-decaying "invisible" part of the propagator and in particular measuring the
Higgs line shape. Strictly speaking one has only fully established the Standard Model,
when one has measured the line shape of the Higgs boson. Since the Standard Model width is
$4\mev$ this is highly challenging.

There are basically two approaches to this problem. One is to build a muon collider.
It appears possible to build a muon collider that would be able to measure the width of the Standard Model
Higgs with moderate precision. However this would leave out the challenge of looking for an invisible, maybe 10\% 
partial Higgs boson at another mass. In order to look for such a signal one needs to radiate off an extra photon,
which reduces the cross-section by a factor of the fine-structure constant. So one would need a muon collider
with a luminosity that is about a thousand times larger than usually considered. This does not look
feasible.

The other option is to build a very large electron-positron collider, where one can measure the Higgs spectrum
from the recoil spectrum in the process $\e^+e^-\rightarrow ZH$, looking only at the outgoing Z-boson~[m9].
This makes extreme demands on the machine. One would have to know the collision energy within $1\mev$.
Also the detectors should be able to measure the outgoing muons with a precision of
$\Delta p/p \approx 10^{-5}$.  A 
very naive rescaling from LEP gives a size of 230 km for the required ring.
Such a ring might (barely) be built, for instance in Fermilab, if the whole
world would work on this in a united way. It is clearly too large for a single region.
For organizational purposes one would need a structure like CERN, but not with countries, but
with regions as units. One could divide the world in 9 regions, that could contribute
according to their level of development: North America, South America, Western Europe, Eastern Europe,
China, India, East-Asia-Pacific, Middle-East, Subsaharan Africa.
Given the present state of the planet and humanity, it is unlikely that humanity will ever build such a machine;
so we will never know for sure whether the simple Standard Model describes the Higgs sector correctly.
The present plans, like FCC and CEPC, are actually not sufficient to settle the question. Moreover we
are interested in having a machine now, not after the LHC. So at present the easiest way forward would be
to go for the ILC in Japan, that is ready to be built. This is clearly not the ultimate machine, but can
settle a number of questions nonetheless. However it should be designed to go to $300\gev$, so one can study
the Higgs propagator up to the 2Z-threshold, below which hadron colliders are not very good.
Except for reasons of national pride, there is no particular reason why other countries could not combine 
to pay for at least half the cost of the ILC.

\section{The principle of global relativity} 

After this discussion we are tempted to conclude, that the NMSM is indeed the \TOU.
So we give in to some hubris and assume that we know the laws of nature well enough.
So we will make a try to get to the next level: is there a way to understand
why this would be the case? Are the laws of nature unique?
 To put it differently: Are there a small number of principles that
would be sufficient to determine the form of the fundamental laws of nature?
Einstein paraphrased the question in the following way: Did God have a choice when He created the world?\\
 
We will not try to answer everything  at the same time, but will consider the theory without the Higgs
boson. Then one is interested in the following only:
Is there a reason for the choice of gauge group and representations?
That is not an unreasonable limitation as a start. After all, before worrying about the 
breaking of a symmetry, it may be necessary to understand the symmetry itself first.
It is actually an old question, going back to the discovery of the muon,
summarized by the famous question of I.~Rabi: who ordered that?\\

What of a handle exists in quantum field theory for such a question?
The only one we know of are anomalies. Since pure gauge theories
with fermions can exist as fundamental theories, when they are asymptotically free,
anomalies in the gauge sector cannot be enough. So we are naturally led towards
the consideration of anomalies involving gravity.
The question of gravitational anomalies has been studied in great detail
in  \cite{alwit,witten}. The main focus of these papers is on anomalies in higher
dimensions, with applications towards Kaluza-Klein theories. Here I will take the view,
that one should maybe go to lower, actually three, dimensions.

In three dimensions it is possible to have, besides the ordinary Yang-Mills or Einstein-Hilbert action,
a parity violating Chern-Simons term in the action \cite{siegel,schonfeld,jackiw1,jackiw2,jackiw3}. 
This gives rise to a parity violating mass term, that is actually quantized
for topological reasons. Radiative corrections give contributions to the Chern-Simons mass and as a consequence 
there are restrictions on the number of fermion fields. Alternatively in the case of massless
fermions there can be a non-perturbative parity anomaly. Most of the discussion in the literature
was focused on the Yang-Mills case. Here also the gauge bosons themselves can renormalize the
Chern-Simons term \cite{sumrob}.
There was some discussion on the three dimensional parity anomaly in  \cite{alwit},
however this part was limited to gauge anomalies. Regarding gravity
it was said: We do not know under what conditions such phenomena occur in general relativity.

Actually, in three dimensions there is not only a parity anomaly, also known
as induced Chern-Simons term, from the fermions, there is also one coming from the
gauge bosons themselves. Subsequently a number of calculations addressed this question
 \cite{indcs1,indcs2,indcs3,indcs4,indcs5}, where only  \cite{indcs1} addressed both the fermion and the vectorboson contribution
to the induced gravitational Chern-Simons term~[m10]. The opposite calculation gives that there is no
contribution to the gauge \CS term from graviton exchange.

Just looking at three-dimensional anomalies will of course not do, since we are living in four dimensions.
However cosmologically speaking this is not so clear. In the simplest scenario, the universe 
basically starts as a point. Going back in time everything shrinks in the same way, but cosmology 
gives more possibilities.\\

Relevant here is  \cite{witten}, which ends with the folllowing sentence:
The choice of S4 corresponds to treating four dimensional space time
as Minkowski space. In the long run, a more delicate choice will be necessary to accomodate cosmological
considerations. It may be that eventually global anomalies will have cosmological applications, 
restricting the large scale topology of space-time.\\

This is one way to look at the problem. If one knows what particle types exist
in the universe, certain conditions on the topology
of spacetime can surely be derived.
However one can also turn the statement around. We could assume, that  "all" topologies are allowed,
also in a cosmological context. Hereby "all" is to be defined in detail,
as we would want to allow for the existence of fermions for example.
Allowing all topologies, but having the same matter content always present in some form,
we can reformulate.
The laws of nature, i.e. the matter content of gauge fields and fermions, should
be the same in all topologies allowed by the Einstein equations.
Formulated this way, the term "principle of global relativity" appears unavoidable.

If this principle is valid it could imply that the possible matter content
is constrained, in the best case being unique.
In order to see what is possible, I will describe a possible cosmological scenario,
that might correspond to the actual universe.  For the principle to apply
it is not strictly necessary that the universe follows the model; it is sufficient
that the model is potentially possible.\\

Observationally there are limits on the topology
of the universe  \cite{observation}. Within the $\Lambda$ cold dark matter model
of the universe the "slab" universe is the least constrained. However real
solid conclusions on the topology of the universe are difficult
to obtain, because of inhomogeneities that grow into galaxy distributions at recent epochs \cite{roukema}. 

The idea is, that one starts with a "slab" universe, where the compact dimension shrinks faster,
going backwards in time, than the other two. At a certain point in time it reaches the Planck scale and disappears.
Then one is in three dimensions and can use the results from the calculations
in  \cite{jochum1,jochum2}. One finds that the dimension of the gauge group must be a multiple of eight; 
SU(5) has three times eight generators. The fermions must come in multiplets of sixteen;
a generation has sixteen fermions. To cancel fermions against bosons one needs precisely
three generations. At first sight this looks a bit like numerology. However it is a rather 
unique possibility within group theory and as far as I know the only argument where something
close to nature comes out. Also the possible symmetry breaking pattern points towards
a breaking pattern leading to the Standard Model. The physics I used is fairly well established
 mathematical physics, that has applications in solid state physics as well. So I did not need to
introduce a large superstructure of new physics, like superstrings in order to derive the results;
it is a conservative approach.

At first sight, it looks strange that one assumes that the universe starts from
a lower dimensional space, but is it really stranger than starting 
the universe from a point, like in normal cosmology?
Or is it stranger than assuming that the present universe started from  
a previous one, as is described by Prof. Penrose in his Nobel prize talk?
In the end it is hard to prove things one way or the other,
as with such transitions one is in the realm, where quantum gravity/geometry
is fundamentally important~[m11,m12].

At least the factor sixteen for the fermions has been found
directly in a four-dimensional context \cite{etxebarria}.
If in a quantum-gravitational context the transition from three to four dimensions could be understood,
one would have a formal derivation, that the Standard Model with three generations is the only possible
low-energy theory. This is of course only valid for the chiral sector, which is why the allowed
class of models is the NMSM, with minimalistic extensions of the Standard Model.

Actually there is work on quantum gravity  \cite{reuter,loll,carlip},
following a more or less canonical approach, that indicates that at large energies
far above the Planck scale, spacetime effectively becomes two dimensional~[m13].
So at low energies spacetime is four dimensional, as we know;
at scales far above the Planck scale it is two dimensional.
So logically speaking, it is natural to assume that at the Planck scale 
spacetime is effectively three dimensional. Altogether one finds this way
a rather consistent picture of the universe including matter fields.
In the geometrical approaches one is mostly focused on gravity and less on 
the matter fields. How to truly unify these may involve new ideas;
the factors eight and sixteen are intriguing. A question is whether we can  find some experimental
verification of the ideas. There is some hint, that there is indeed a preferred direction
in the universe, that might be related to a compactified dimension, but there is little
possibility for a definite conclusion, since one is always affected by the question of cosmic 
variance. More promising is therefore to look at what the prediction of unification
into SU(5) tells us.\\

\section{SU(5) and dark matter}
We have now shown, with all caveats, that the gauge group of nature must be SU(5).
This immediately raises a problem. It is well known that the Standard Model
particle content cannot lead to a unification of forces, when one lets the  
coupling constants run as a function of the scale. It is known that supersymmetry
can change this, but supersymmetry is not seen and is not in the class
of minimalistic extensions. So one has to start with different additions to 
the model. The easiest way is to introduce a ${\bf 24}$ of fermions of SU(5).
Introducing one Majorana ${\bf 24}$ helps with unification, but is not really
successful. This also does not satisfy the condition that the number of fermions
must be a multiple of 16. So the easiest way out is to start with 
a Dirac ${\bf 24}$ \cite{su5}. Then unification is easy, because different patterns of mass splitting
are possible after symmetry breaking. While a Dirac fermion ${\bf 24}$  appears the most
natural, any combination of an even number of real ${\bf 24}$ representations would satisfy the
16-fold condition. With these extra fermions the scale of unification can be
varied over a wide range. The scale can be such that proton decay is out of range
or just around the corner for proton decay experiments.

Another feature is that after symmetry breaking a ${\bf 24}$ leads to triplet fermions.
These are candidates for the dark matter of the universe \cite{cirelli}.
Actually at the tree level the charged and the neutral fermions have the same mass,
but through radiative corrections the charged one is 165 MeV heavier.
It decays into the neutral one with emission of a charged pion.
This makes this type of dark matter particularly hard, probably impossible
to see at the LHC. The signal would be a soft pion with missing energy.
WIMP triplet fermions are in a sense preferred, because they do not couple
directly to ordinary matter via Z or Higgs exchange, but only
through loops, so their cross-section is much smaller.
WIMP dark matter with direct exchange is strongly constrained by
experiment  \cite{lindner}. The triplets are more constrained through indirect limits,
meaning dark matter annihilation leading to $\gamma$ rays. Having a single Dirac fermion
appears to be ruled out by the
 HESS experiment, the reason being a large Sommerfeld
enhancement in the annihilation cross section, precisely in the predicted mass range,
needed to explain dark matter. This can be cured by having two Dirac triplets,
or more general an even numer of real triplets \cite{cai}. Also the role of the singlet in
the ${\bf 24}$ is not quite clear. The new information, that apparently
a large part of dark matter consists of black holes, may allow for more possibilities.
The \v Cerenkov telescope array (CTA) should be able to clarify the situation.

\section{Not the theory of everything}
The idea of a theory of everything  is of course not to describe everything
that happens, for instance all craziness in human affairs, in detail. It is known,
that even for simple systems this is not possible, because of the phenomenon of chaotic behaviour.
The idea is to derive the fundamental rules from simple principles and in particular to calculate 
the constants of nature from the theory. Even though, as we have seen above, we have some idea of the general
structure of the world, all attempts to calculate coupling constants and masses in the literature
have been a miserable failure. Also all attempts to quantize gravity have been less than successful.
Both these questions are not addressed in the theory above. Both these questions have also been discussed in thousands 
of papers. I will argue that these questions lie beyond the range of physics, understood as an experimental science.

The theory described above is a good theory to describe the universe as we know it, but cannot explain the coupling constants and masses.
Actually in a way the theory appears to be too good. Calculations show that for instance the Higgs potential
becomes essentially flat precisely near the Planck scale, which surely means something. This feature would be spoiled if there
were a mechanism, whereby the masses of the particles are determined by new dynamics below the Planck scale.  
Of course we do not know of such a mechanism anyway, but the argument shows, that high energy colliders cannot
be expected to see anything interesting in flavour dynamics. Therefore any dynamics that determines the coupling
constants and masses must come from physics beyond the Planck scale. Is there any way we could conceivably probe
this region of nature? The answer appears to be a resounding NO. In inflationary theories, even when one
enhances quantum gravitational effects with non-minimal couplings, these quantum gravitational effects have far too small
an impact to be seen in observable quantities like density fluctuations \cite{steinwachs}.  

So what can we do? Astronomy and physics can test the precise form the \TOU will take.
As mentioned above, a number of experiments and observations are underway or could in principle be 
performed, though they would be quite expensive. Regarding the \TOE one can only hope to find a deeper 
mathematical principle, that is sufficiently strong to determine the dynamics in a unique way.
However history tells us that, without experimental guidance, this could well be a fool's errand~[m14].\\

{\bf Acknowledgements}
This work would not have been possible without Tini, who was like a father to me.
Secondly I am thankful to my gymnasium teacher Jan van der Putten, who introduced relativity to me
and to my Drs.-thesis advisor Gerard 't Hooft, for teaching me about topology
in field theory. Further thanks are due to my 54 collaborators
on scientific papers and the roughly 600-800 people with whom I worked on reports or had
fruitful scientific discussions. \\
\vskip 1.2cm

{\Large \bf Memories, history and remarks}
\begin{enumerate}[label={[m\arabic*]}]

\item On Tini's 80th birthday I gave a talk about Higgs physics \cite{tini80}.
Tini told me he had rather have heard something about SU(5).
So I prepared this talk for his 90th birthday. I have discussed
the subjects in this paper with him. Normally he would find faults in
my ideas, but this time he only found it a pity that these theories could not be 
measured. This is not really true, but it will be quite expensive. The terminology
"principle of global relativity" was first used in a discussion during
my talk at the AEI in Potsdam on 26th January 2018. In writing it first appeared 
at the 15th Patras meeting in Freiburg on 3rd June 2019.

\item Actually MNSM, which was originally a typing error, 
  might be a more interesting name for the theory.
  It would stand for Minimalistic Non-Standard Model.

\item Tini himself was not directly involved in calculations for the production
of particles at the LHC. But SCHOONSCHIP was used in the first calculation of
boosted Higgs production  \cite{ellis}.
Moreover from 1987-1989 there were two groups, systematically
calculating the production of weakly interacting particle pairs from gluon fusion.
One was in CERN, consisting of E.W.N. Glover and myself.
We used SCHOONSCHIP on an Atari computer, to express the amplitudes
in irreducible scalar integrals. The others were D.A. Dicus and Chung-Kao,
who used Tini's FORMF program, that contained numerical routines also for
tensor integrals. An example is Higgs pair production  \cite{glover,chungkao}.

\item The idea that there would be strong interactions or that there would be more than just the 
simple Standard Model, was rather wide-spread at the time. For instance I had an early
discussion with G. 't Hooft:
\noindent G.: {\em   How can the Higgs be so light? There is no symmetry to protect its mass!}\\
J.: {\em   Scale invariance?}\\
G.:   {\em Yes, but that is not a quantum symmetry!}\\
J.: {\em   OK, how about supersymmetry?}\\
G.: {\em  Yes, but that does not exist!}\\
J.: {\em   So what could it be?}\\
G.: {\em  It has to be like with the strong interactions, but a bit different.}

\noindent This might be true. The Higgs field in the HEIDI models is much like
the $\sigma$-field in QCD, but more weakly interacting. Also scale invariance
may still play a role somewhere and supersymmetry appears not to exist.

\item Traditionally, doctoral theses in the Netherlands are printed as a small book.
As my thesis defence was in Utrecht, but my work had been done in Ann Arbor and there were some timing
problems, my booklet was printed in Ann Arbor. For the local printers this was an interesting experience
and I got a tour of the factory. Anyway, because of this, the thesis was imported into the Netherlands
and I had to pay import duty on it. Tini was quite indignant about this: {\em Jochum, that is all
wrong, they should have called me and I would have told them that it is not worth anything}.

\item Working together with Tini could be an interesting experience.
As the two-loop calculation we were doing was quite difficult for that time,
things were a bit tense sometimes. He once threw me out of his office, which with other
students and supervisors could be problematic. I just found it curious.
We had gotten side-tracked in a discussion on the meaning of the renormalization
of the Faddeev-Popov ghost fields. I interpreted this as a redefinition of the basis
in the Lie algebra of the gauge-symmetry. As the ghost fields do not contribute to the
order we calculated, this may have sounded smart, but was besides the point.
Near the end of the calculation, things did not fit.
Tini said, that he was at the end of his Latin. As I have a classical gymnasium education,
I answered: {\em No problem, we continue in Greek}. The bug was quickly found,
we had to expand some integrals a bit further.

\item Alfred Hill was a German physicist who studied in Groningen and did his doctorate with Tini
in Michigan. Besides the Hill-model, which was his thesis, he wrote an unpublished 
 paper on higher-dimensional anomalies. His Dutch was so perfect, that from this
you could tell he was not really Dutch. He died in Lockerbie.

\item Tini and I often discussed the question what accelerators should be built.
He has often told me that he had considered the energy of LEP to be too small. He was right there;
a bit more would have allowed LEP to discover the Higgs boson and  study it in precise detail.
But of course there were engineering limitations and no one knew where the Higgs mass would be.
Also there was for a long time a fairly general expectation, that the Higgs would be heavy.
For a very heavy Higgs boson only indirect effects would be visible, for instance in 
4W couplings. Tini and I were together in DESY in 1990. We there often discussed the possibility
of building a linear collider. 
For an energy we came up with $500\gev$. The precise reason for this energy is a bit unclear,
but at this scale 4W couplings become measurably different from the Standard Model due
to radiative corrections from a heavy Higgs boson. Also one comes in the region where one can
study triple vector boson production. Anyway $500\gev$ became the standard, with an option to go to
$800\gev$. For these energies circular colliders are no option.

\item I first considered the option of a large circular  $e^+e^-$ collider as
a Higgs factory in 2009. This was as a consequence of my studies on the possibility of
 not seeing the Higgs boson at the LHC,
which I presented in Moriond 2007 and 2008 and Blois 2009.
I then mentioned the possibility at the LC forum in DESY Hamburg 14 June 2010
and M\"unchen 14 Juli 2011. However here the focus was more on a $300\gev$ linear collider.
At the time there was no design for a $300\gev$ collider; all was concentrated on $500\gev$.
I had some correspondence with Prof. Brian Foster on the possibility of lowering the energy.
I also mentioned the possibility at the Veltman 80th birthday meeting in NIKHEF 24 June 2011.
Things became more official after the Moriond meeting March 3-10 2012,
where there were also contributions by experimentalists from CERN.
The considerations became serious after the European Strategy meeting in Krak\'ow September 2012,
where Nigel Glover and I also sent in a contribution.

\item Sumathi Rao was my officemate in Fermilab. After she and Rob Pisarski had finished 
their Chern-Simons Yang-Mills calculation \cite{sumrob}, we got into a discussion about the gravitational case.
Rob thought this would be too difficult. However I had experience with SCHOONSCHIP,
Tini's algebraic manipulation program, so I knew it would not be
too bad. I also used Tini's lectures on gravity from the 1975 les Houches
school for the vertices. 

\item Of course Tini was a pioneer in quantum gravity calculations  \cite{vandam,instpoin}.
While gravity involves the manipulation of many indices, Tini never left the underlying physics
out of sight.
Typical is the discussion he had with M. Gell-Mann. They were discussing about quarks and
Gell-Mann said: {\em Tini, these are just indices}, upon which Tini started to jump
up and down: {\em This is just indices?}

\item When I started as a doctoral student with Tini, I was of course full of optimism,
convinced that we would soon solve quantum gravity and everything else.
I even wrote a small paper \cite{ng}. Tini, who had been around a bit longer, told me:
{\em Jochum, quantum gravity is a black hole; when you jump in, you will never get out!}
For me at the time, that was a wise advice and I did not jump in.
This in contrast to my former student Eugen Radu \cite{radu}.

\item Tini used poles in $d=2$ as a gauge invariant way to define
quadratic divergences  \cite{infraredultraviolet}. I try here a somewhat more dynamical
approach to dimensional reduction. Some modern approaches towards quantum gravity
point in this direction as well.  \cite{reuter,loll,carlip}. 

\item The last subjects I discussed with Tini, not long before he died, were some ideas along these lines.
And of course he told me they were crazy. He was probably right, but maybe not; we will see or maybe not.

\end{enumerate}

\end{document}